\documentclass[aps,pra,showpacs,twocolumn,superscriptaddress]{revtex4}

\usepackage{graphicx,subfigure,color}
\usepackage{latexsym,amsmath,amssymb,amsfonts,amsthm,enumerate}
\usepackage{amscd}
\usepackage{bm}

\begin{document}
\bibliographystyle{apsrev}

\title{Heralded noiseless amplification and attenuation of non-gaussian states of light}

\author{C. N. Gagatsos}
\affiliation{Quantum Information and Communication, Ecole polytechnique de Bruxelles,
Universit\'{e} libre de Bruxelles, 1050 Brussels, Belgium}
\author{J. Fiur\'{a}\v{s}ek}
\affiliation{Department of Optics, Palack\'{y} University, 17. listopadu 1192/12, CZ-771 46 Olomouc, Czech Republic}
\author{A. Zavatta}
\affiliation{Istituto Nazionale di Ottica, INO-CNR, Largo Enrico Fermi, 6, I-50125 Firenze, Italy and
Department of Physics and LENS, University of Firenze, 50019 Sesto Fiorentino, Firenze, Italy}
\author{M. Bellini}
\affiliation{Istituto Nazionale di Ottica, INO-CNR, Largo Enrico Fermi, 6, I-50125 Firenze, Italy and
Department of Physics and LENS, University of Firenze, 50019 Sesto Fiorentino, Firenze, Italy}
\author{N. J. Cerf}
\affiliation{Quantum Information and Communication, Ecole polytechnique de Bruxelles,
Universit\'{e} libre de Bruxelles, 1050 Brussels, Belgium}
\date{\today}

\begin{abstract}
We examine the behavior of non-Gaussian states of light under the action of probabilistic noiseless amplification 
and attenuation. Surprisingly, we find that the mean field amplitude may decrease in the process of noiseless amplification -- or increase in the process of noiseless attenuation, a counterintuitive effect that Gaussian states cannot exhibit. This striking phenomenon could be tested with experimentally accessible non-Gaussian states, such as single-photon added coherent states. We propose an experimental scheme, which is robust with respect to the major experimental imperfections such as inefficient single-photon detection and imperfect photon addition. In particular, we argue that the observation of mean field amplification by noiseless attenuation should be feasible with current technology.
\end{abstract}

\maketitle

\section{Introduction}

In quantum optics, it has long been known that the amplification of light unavoidably comes with noise, due for instance to spontaneous emission in parametric down conversion \cite{Caves82}. This causes a fundamental problem in quantum communication, for noise is generally harmful to quantum information (e.g., it sets a limit on the amount of secret key bits that can be shared between two parties in quantum key distribution, see \cite{Scarani09}). Recently, however, it has been realized that the amplification process can, in principle, be made noiseless if one turns to a probabilistic (heralded) process instead of a deterministic process. Specifically, a \emph{heralded noiseless amplifier} can be devised, which brings the added noise variance arbitrarily close to zero at the price of a vanishing success probability \cite{Ralph08}. Soon after it had been proposed, this concept of heralded noiseless amplification was successfully demonstrated in the laboratory by several teams \cite{Xiang10,Ferreyrol10,Usuga10,Zavatta11}. In these experiments, the input state is typically a superposition of the vacuum and single-photon state, and the noiseless amplifier is shown to enhance the single-photon amplitude in this superposition. One can also consider a dual process, called \emph{heralded noiseless attenuation} \cite{Micuda}. Applied to a superposition of the vacuum and single-photon state, it decreases the single-photon amplitude without adding noise. In other experiments, the noiseless amplification of a polarization qubit was also demonstrated using two noiseless amplifiers, one for each polarization mode \cite{Osorio12,Kocsis13}.

Since the noiseless amplifier enhances the intensity of a light state without adding noise, it is naturally a good candidate to improve the performances of quantum communication or metrology protocols. Its potential applications include, for instance, continuous-variable quantum error correction \cite{Ralph11} or phase-insensitive single-mode squeezing \cite{GKC12}. It appears especially useful in the context of quantum key distribution, as it has been theoretically shown to improve the key rate of device-independent discrete-variable protocols \cite{Gisin10,Curty11} as well as the range and tolerable excess noise of continuous-variable Gaussian protocols \cite{Blandino12,gausspostselect,Ralph13}. Interestingly, the combination of a noiseless attenuator and amplifier at the two ends of a communication line provides a means to reduce the line losses without adding noise \cite{Micuda}, which can be exploited in quantum key distribution (note that the noiseless amplifier or attenuator does not necessarily need to be realized in practice, but can be emulated via a postselection procedure \cite{gausspostselect}). 

In this paper, we investigate in depth the action of these noiseless transformations on arbitrary states of light. Since by construction the noiseless amplifier enhances the mean field amplitude (i.e., the mean value of the annihilation operator) of a coherent state, it has implicitly been assumed that this behavior is universal. Surprisingly, we show here that it is not necessarily the case when non-Gaussian states of light are considered. Despite the fact that the mean photon number always increases via noiseless amplification (or decreases via noiseless attenuation), we observe that the transformation of the mean field amplitude is more subtle. After a brief presentation of noiseless amplification (attenuation) in Section~II, we derive in Section~III the transformation of the Husimi Q-function that it effects, which allows us to prove that the mean field amplitude of any Gaussian state can only increase via noiseless amplification (or decrease via noiseless attenuation), in accordance with our intuition. Then, in Section~IV, we analyze the counterintuitive effect of amplitude reduction by noiseless amplification (or amplitude enhancement by noiseless attenuation) that can be exhibited by certain non-Gaussian states. We provide examples of pure and mixed non-Gaussian states where this striking effect is visible. Finally, Section~V is devoted to the proposal of an experimental scheme which could be used to demonstrate that the mean field amplitude of a single-photon added coherent state is increased in the process of noiseless attenuation. While taking into account the inefficiency of the single-photon detector and an imperfect source, we show that this scheme could be accessible with current technology. Our conclusions are given in Section~VI.

\section{Noiseless amplification and attenuation}

The noiseless amplifier probabilistically enhances the amplitude of a coherent state as
\begin{equation}
|\alpha\rangle  \to |g\alpha\rangle
\end{equation}
where $g>1$ is the amplitude gain. It can be described by a quantum filter $F$ (a trace-decreasing CP map with a single Kraus operator $F$) such that
\begin{equation}
\rho  \to F \rho F^\dagger
\end{equation}
where the filter $F=c g^{\hat{n}}$ is diagonal in the Fock state basis $|n\rangle$ and $c$ is a real constant. The trace non-increasing condition $F^\dagger F \le \openone$ implies that $|c|^2 g^{2n}\le 1$, $\forall n$, which is possible only if $c=0$; hence, the success probability $\mathrm{Tr}(F \rho F^\dagger)$ of this \emph{ideal} noiseless amplifier vanishes. Mathematically, this is because the operator $g^{\hat{n}}$ is unbounded for $g>1$. However, a  \emph{non-ideal} version of the noiseless amplifier can be defined by truncating the Fock state basis at $|N\rangle$. Then, the trace non-increasing condition is fulfilled provided $|c|^2 g^{2N} = 1$; hence, the success probability scales as $g^{-2N}$ and can be made strictly larger than zero as long as $N$ is finite. In other words, a noiseless amplifier can be implemented with non-zero success probability only within a finite-dimensional subspace of the Fock space.  We will ignore this subtlety in the rest of this paper, and consider the ideal noiseless amplifier that is simply associated with the quantum filter $g^{\hat{n}}$.

\begin{figure}[!h!]
\centering
 \includegraphics[width=0.8\linewidth]{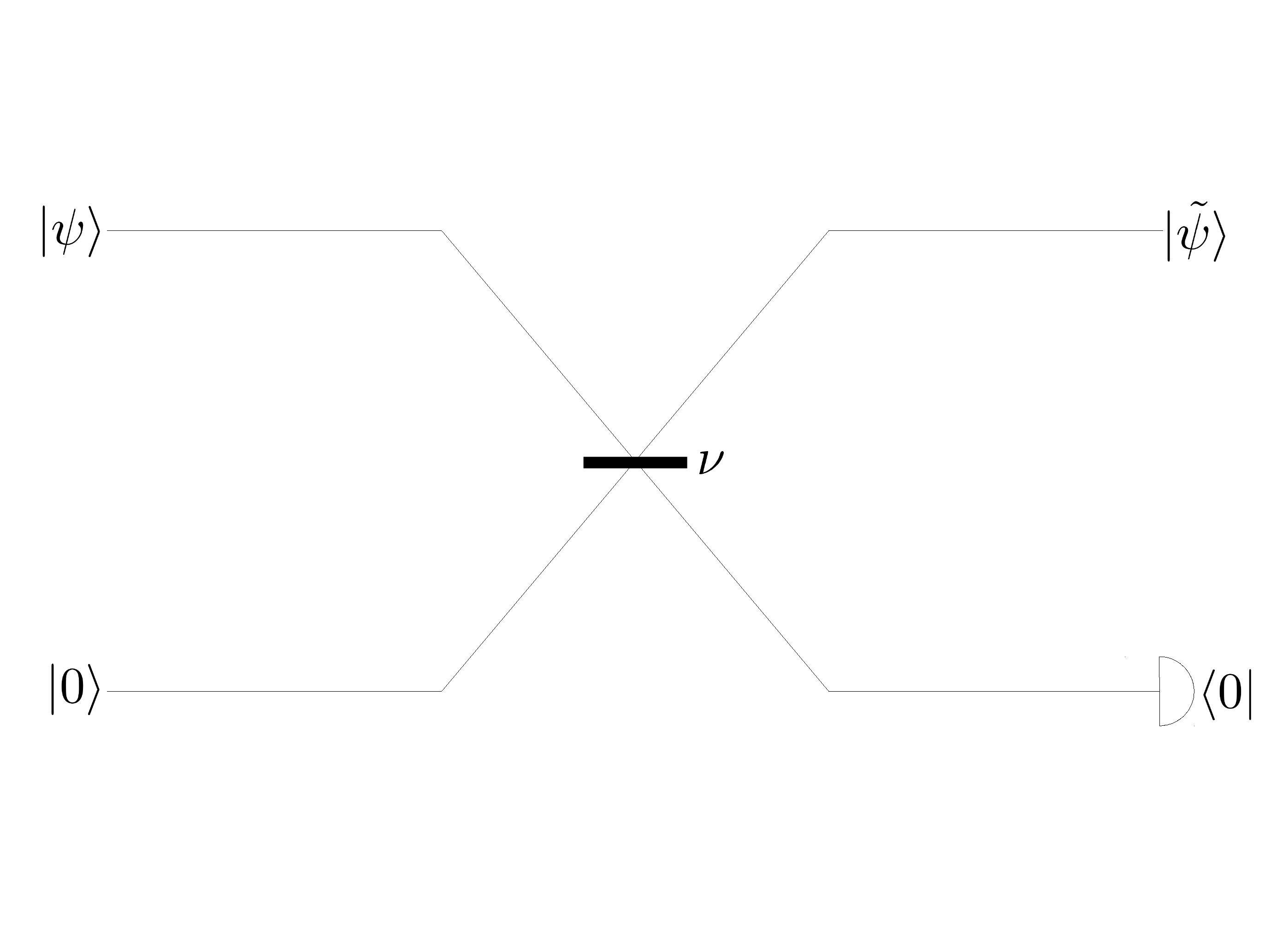}
\caption{Noiseless attenuator. In a beam splitter with amplitude reflectance $\nu$ the lower input mode is set to vacuum and we post-select on vacuum in the lower output mode.}
\label{att}
\end{figure}

Noiseless attenuation corresponds to the same quantum filter, but taking $g=\nu<1$. In contrast with noiseless amplification, it corresponds to a bounded operator $\nu^{\hat{n}}$ for $\nu<1$, so it can be implemented exactly with a success probability that is strictly larger than zero. Indeed, the quantum filter $\nu^{\hat{n}}$ can be realized, for instance, by processing the input state through a beam splitter of amplitude reflectance $\nu$ whose auxiliary input port is prepared in the vacuum state $|0\rangle$, and then conditioning on projecting the state of the auxiliary output port onto the vacuum state $|0\rangle$, as shown in Fig. \ref{att} \cite{Micuda} .
 
It is easy to see that for an input state $|\psi\rangle=\sum_n c_n |n\rangle$, with $\sum_n |c_n|^2=1$, the final state will read
\begin{eqnarray}
 |\tilde \psi \rangle \propto \sum_n \nu^n \, c_n |n\rangle.
 \label{finalstateatt}
\end{eqnarray}
Intuitively, we understand that the heralded filtering operation preferentially keeps low-$n$ Fock states since $\nu^n$ exponentially decays with $n$,
so in this sense the state is attenuated. Conversely, if we formally consider amplitude reflectance larger than $1$, we will get an output state which can be
interpreted as a noiselessly amplified state, where large-$n$ Fock states are preferentially post-selected. This formal equivalence allows us to analyze the effect of both conditional operations 
simultaneously. 

Let us clarify the intuition behind saying that $g^n$ amplifies the state, or $\nu^n$ attenuates the state. It so happens that this intuition holds true as far as the mean photon number  $\langle\hat{n}\rangle$ is concerned, but may be contradicted if we probe the mean field amplitude $\langle \hat a \rangle$ of certain non-Gaussian states (cfr. Section~III). 
As a first step, we will prove here that $\langle\hat{n}\rangle$ is necessarily increased (decreased) under the action of noiseless amplifier $g^n$ (attenuator $\nu^n$).
For simplicity, we consider single-mode states, but the argument can be extended to multimode states. An arbitrary input state $\rho$ can be expressed in Fock basis as
\begin{eqnarray}
\rho=\sum_{n,m=0}^\infty \rho_{mn} | n \rangle \langle m|.
\label{inFock}
\end{eqnarray}
where  $\rho\ge 0$ and $\sum_{n=0}^\infty \rho_{nn} =1$.
The amplified (attenuated) state is
\begin{eqnarray}
\tilde{\rho}=\frac{\sum_{n,m=0}^\infty g^{n+m} \rho_{mn}|n\rangle\langle m|}{\sum_{n=0}^{\infty} g^{2n} \rho_{nn}}
\label{finalRho}
\end{eqnarray}
and its mean photon number is given by
\begin{eqnarray}
 \langle \tilde{n} \rangle= \frac{\sum_{n=0}^{\infty} n g^{2 n} \rho_{nn}} {\sum_{n=0}^{\infty} g^{2n} \rho_{nn}}
 \label{meanPhotons}
\end{eqnarray}
We shall assume that this state is physical, i.e., the sum in the denominator exists and has a finite value (this is not necessarily true for some input states and $g>1$).
The derivative of Eq. (\ref{meanPhotons}) with respect to $g$ is
\begin{eqnarray}
 \frac{d \langle \tilde{n} \rangle}{dg}=\frac{1}{N^2}\sum_{m,n=0}^{\infty} n(n-m) e_{nm},
 \label{meanPhotonsDer1}
\end{eqnarray}
where
\begin{eqnarray}
N=\sum_{n=0}^{\infty} g^{2n} \rho_{nn}
\end{eqnarray}
and
\begin{eqnarray}
e_{nm}=e_{mn}=2g^{2(n+m)-1}\rho_{nn}\rho_{mm} \geq 0.
\end{eqnarray}
Equation (\ref{meanPhotonsDer1}) can be rewritten as
\begin{eqnarray}
  \frac{d \langle \tilde{n} \rangle}{dg}=\frac{1}{N^2}\sum_{n=0}^{\infty}\sum_{m=0}^{n} (n-m)^2 e_{nm}
 \label{meanPhotonsDer2}
\end{eqnarray}
from which we conclude that
\begin{eqnarray}
 \frac{d \langle \tilde{n} \rangle}{dg} \geq 0.
\end{eqnarray}
Thus, the mean photon number of any physical state increases when $g$ increases (amplification), or decreases when $g$ decreases
(attenuation). If the input state is a Fock state, which is an eigenstate of the operator $g^{\hat{n}}$, then the mean photon number remains constant under noiseless amplification or attenuation.

\section{Noiseless transformation of Gaussian states}

Let us now discuss how noiseless amplification (attenuation) transforms Gaussian states, paying a particular attention to the properties of the mean field. Let us begin by recalling that the operator $g^{\hat n}$ transforms a coherent state $|\alpha\rangle$ as
\begin{eqnarray}
g^{\hat{n}}|\alpha\rangle= e^{(g^2-1)|\alpha|^2/2}|g\alpha\rangle.
\label{hnlaDef}
\end{eqnarray}
where $\alpha$ is a complex number. This transformation suggests to decompose any input state into
the overcomplete basis of coherent states. Then, one has to evolve every component coherent state according to the transformation (\ref{hnlaDef}). 
A natural idea may be to use the Glauber-Sudarshan $P$ representation of the input state, that is
\begin{eqnarray}
\rho=\int d^2\alpha \, P(\alpha) |\alpha\rangle\langle \alpha|.
\label{Prepr}
\end{eqnarray}
If we apply the transformation of Eq. (\ref{hnlaDef}) to both sides of Eq. (\ref{Prepr}), we obtain the $P$ representation of the transformed state $\tilde \rho \propto g^{\hat n} \rho g^{\hat n}$, namely \cite{BlandinoPhD}
\begin{eqnarray}
\tilde{P}(\alpha)\propto e^{(1-1/g^2)|\alpha|^2} P\Big(\frac{\alpha}{g}\Big).
\label{PreprAmp}
\end{eqnarray}
where the symbol $\propto$ indicates that $\tilde{P}$ needs to be normalized. 
In the case at hand, we find it more elegant to consider instead the Husimi $Q$-function. For an 
arbitrary quantum state $\rho$, the $Q$-function is defined as
\begin{equation}
Q(\alpha)=\frac{1}{\pi} \langle \alpha |\rho|\alpha\rangle,
\end{equation}
For a Gaussian state with covariance matrix $\gamma$
and vector of mean values of quadrature operators $d=(\langle \hat{x}\rangle,\langle \hat{p}\rangle)^T$, it can be expressed as
\begin{equation}
Q(\alpha)= \frac{2}{\pi \sqrt{\det(\gamma+I)}} \exp[-(r-d)^T \Gamma (r-d)].
\label{gaussQ}
\end{equation}
Here $\Gamma=(\gamma+I)^{-1}$, $I$ denotes the identity matrix, and $r=(\sqrt{2}\alpha_R,\sqrt{2}\alpha_I)^T$,  where $\alpha_R$ and $\alpha_{I}$
denote the real and imaginary parts of $\alpha$. We use the normalization convention where the covariance matrix of the vacuum is equal to the identity matrix, 
while the variance of the vacuum quadratures reads $1/2$.

We recall that $Q(\alpha)$ can be viewed as the probability density for the complex outcome $\alpha$ of a heterodyne measurement performed on state $\rho$, which consists in projecting onto the coherent-state basis. It is then possible to back-propagate each coherent state through the noiseless amplifier or attenuator, as done in Ref. \cite{gausspostselect}.
The $Q$-function $\tilde{Q}(\alpha)$ of the transformed state $\tilde{\rho} \propto g^{\hat{n}} \rho g^{\hat{n}}$ can then be written as 
\begin{equation} 
\tilde{Q}(\alpha) \propto \frac{1}{\pi} \langle \alpha| g^{\hat{n}}  \hat{\rho} g^{\hat{n}} |\alpha\rangle = e^{(g^2-1)|\alpha|^2} Q(g\alpha),
\label{transfQ}
\end{equation}
where we have used Eq. (\ref{hnlaDef}) and the symbol $\propto$ indicates that $\tilde{Q}$ needs to be normalized.
Note that if $Q(\alpha)$ is expressed as in Eq. (\ref{gaussQ}), its Gaussian form is preserved by transformation (\ref{transfQ}), so that Gaussian input states are mapped onto Gaussian output states. 
In particular, the corresponding transformations on $\gamma$ and $d$ can be determined by looking
at the exponent in $\tilde{Q}(\alpha)$. After some algebra, we find
\begin{equation}
\tilde{\Gamma}= g^2\Gamma-\frac{g^2-1}{2} I
\end{equation}
which implies that the covariance matrix transforms as
\begin{equation}
\tilde{\gamma}=\left[g^2(\gamma+I)^{-1}-\frac{g^2-1}{2} I\right]^{-1}-I.
\label{CMtilde}
\end{equation}
Similarly, the vector of mean values transforms as
\begin{equation}
\tilde{d}= 2g\left[(g^2+1)I-(g^2-1)\gamma)\right]^{-1}d.
\label{dformula}
\end{equation}
Here, tilde denotes the parameters of the Gaussian state after noiseless amplification or attenuation. 
Note that Eq.~(\ref{dformula}) agrees with a formula for the displacement of a noiselessly amplified Gaussian state derived in Ref. \cite{walk13}.

We can also easily check that Eqs. (\ref{CMtilde}) and (\ref{dformula}) are consistent with the formulas obtained in Ref. \cite{GKC12} for the noiseless amplification of a squeezed state of light, whose covariance matrix is given by
\begin{equation}
\gamma= \left(
\begin{array}{cc}
{\rm e}^{-2s} & 0 \\
0 & {\rm e}^{2s} 
\end{array}
\right)
\end{equation}
with $s$ being the squeezing parameter. We have
\begin{equation}
\Gamma= (\gamma+I)^{-1} = \left(
\begin{array}{cc}
\frac{1+\tanh s}{2} & 0 \\
0 & \frac{1-\tanh s}{2} 
\end{array}
\right)
\end{equation}
implying
\begin{equation}
\tilde \Gamma=  \left(
\begin{array}{cc}
\frac{1+g^2 \tanh s}{2} & 0 \\
0 & \frac{1- g^2 \tanh s}{2} 
\end{array}
\right)
\end{equation}
so we conclude that the covariance matrix of the amplified state is that of another squeezed state
\begin{equation}
\tilde\gamma= \left(
\begin{array}{cc}
{\rm e}^{-2s'} & 0 \\
0 & {\rm e}^{2s'} 
\end{array}
\right)
\end{equation}
with stronger squeezing (the output squeezing parameter $s'$ satisfies $\tanh s' = g^2 \tanh s$). The transformation of the vector of mean values, Eq.~(\ref{dformula}), gives
\begin{equation}
\left(
\begin{array}{c}
 \langle \tilde{x}\rangle \\
 \langle \tilde{p}\rangle
\end{array}
\right)
= g \left(
\begin{array}{cc}
\frac{1+\tanh s}{1+\tanh s'} & 0 \\
0 & \frac{1-\tanh s}{1-\tanh s'}
\end{array}
\right)
\left(
\begin{array}{c}
 \langle x \rangle \\
 \langle p \rangle
\end{array}
\right)
\end{equation}
in perfect agreement with Ref. \cite{GKC12}.

Now, we treat the case of an arbitrary Gaussian input state. Since the operator $g^{\hat{n}}$
commutes with a unitary phase shift $e^{i\phi \hat{n}}$, we can without loss of generality assume that the covariance matrix is diagonal,
\begin{equation}
\gamma= \left(
\begin{array}{cc}
2V_x & 0 \\
0 & 2V_p 
\end{array}
\right),
\label{CMdiag}
\end{equation}
where $V_x$ and $V_p$ denote the variances of amplitude and phase quadratures, respectively, which obey the Heisenberg uncertainty
relation $V_x V_p \geq \frac{1}{4}$. If we insert the diagonal covariance matrix (\ref{CMdiag}) into Eq. (\ref{dformula}), we get
\begin{equation}
\tilde{d}_j= \frac{2g}{(1+g^2) + 2V_j (1-g^2)} d_j,
\end{equation}
where $j=x,p$. The effective amplification gain is thus different for the amplitude and phase quadratures, and it depends on the
variance $V_j$ of the quadrature $j$, namely
\begin{equation}
G_{\mathrm{eff},j} \equiv {\tilde{d}_j \over d_j} = \frac{2g}{(1+g^2)+2V_j(1-g^2)}.
\label{eq-Geff}
\end{equation}
This can be rewritten as
\begin{equation}
\frac {G_{\mathrm{eff},j} - g}{G_{\mathrm{eff},j}} = (g^2-1)(V_j-1/2)
\end{equation}
so that for the noiseless amplifier ($g>1$) we have
\begin{equation}
G_{\mathrm{eff}}^{V_j<1/2}< g < G_{\mathrm{eff}}^{V_j>1/2}
\end{equation}
while for the noiseless attenuator ($g<1$) we have
\begin{equation}
G_{\mathrm{eff}}^{V_j>1/2}< g < G_{\mathrm{eff}}^{V_j<1/2}
\end{equation}
In other words, in both cases, the effective gain is sublinear for the squeezed quadrature ($V<1/2$) and superlinear for the antisqueezed quadrature ($V>1/2$). Of course, in between these cases, we have simply a linear effective gain $G_{\mathrm{eff}}^{V_j=1/2}=g$. At this point, we note that the squeezed quadrature with the lowest possible variance must be considered in order to find the minimum effective gain that the noiseless amplifier may exhibit, as well as the maximum effective gain that the noiseless attenuator may exhibit.

Let us prove that noiseless amplification always increases the amplitude of Gaussian states. Remember that noiseless amplification may generate unphysical states from certain input Gaussian states. The amplified state is physical iff the output covariance matrix is positive definite, which is equivalent to the matrix inequalities $I>g^2\Gamma+(1-g^2)I/2 >0$.
Both these inequalities are  equivalent to
\begin{equation}
\underset{j}{\max}(V_j)< \frac{1}{2}\frac{g^2+1}{g^2-1}.
\end{equation}
and the denominator of Eq. (\ref{eq-Geff}) vanishes if the variance $V_j$ reaches this upper bound, making $G_{\mathrm{eff},j}$ diverge.
Taking this constraint into account, the squeezed quadrature variance of the input state is lower bounded by
\begin{equation}
\underset{j}{\min}(V_j) > \frac{1}{2}\frac{g^2-1}{g^2+1}
\end{equation}
which, when plugged into Eq. (\ref{eq-Geff}), gives
\begin{equation}
G_{\mathrm{eff},j} > \frac{1+g^2}{2g} > 1
\end{equation}
Thus, when $g>1$, one finds that $G_{\mathrm{eff},j}> 1$ for all input states leading to physical output states.

Similarly, by considering the case $g<1$, one can prove that the noiseless attenuation always reduces the amplitude of Gaussian states
and $G_{\mathrm{eff},j}<1$. In this latter case, there is no physical constraint on the admissible input states because noiseless attenuation
is a physically allowed operation that can be implemented with finite success probability on any input state.
Thus, the variance of the squeezed quadrature is simply lower bounded by $\underset{j}{\min}(V_j) > 0$.
When plugging this into Eq. (\ref{eq-Geff}), we obtain
\begin{equation}
G_{\mathrm{eff},j} < \frac{2g}{1+g^2} < 1
\end{equation}
for all input states.

\section{Noiseless transformation of non-Gaussian states}

The heralded noiseless amplifier (attenuator) is a transformation that increases (decreases) the complex amplitude $\alpha$ of a coherent state $ |\alpha\rangle$ without adding noise, that is $|\alpha\rangle \rightarrow |g\alpha\rangle$. We have proven, in the previous Section, that the same behavior holds true for the mean amplitude of any (possibly mixed) Gaussian state.
One could therefore naively expect that this remains true for all states. Surprisingly, we will show 
that the mean amplitude of a non-Gaussian state  can actually be attenuated by noiseless amplification, or amplified by noiseless attenuation.
We first illustrate this counterintuitive effect on two simple and instructive examples of states that can be expressed
as superpositions of a finite number of Fock states. In a third example, a non-Gaussian mixed state will also be shown to exhibit this effect.
In the next Section, we will design a scheme for experimentally verifying
the mean field amplification by noiseless attenuation that is robust against most experimental imperfections.

As a first example, let us consider the superposition of vacuum and single-photon state,
\begin{equation}
|\Psi_1\rangle=c_0|0\rangle +c_1|1\rangle,
\label{psi1}
\end{equation}
where without loss of any generality we assume that $c_0$ and $c_1$ are real and $c_0^2+c_1^2=1$.
The coherent amplitude of this state then reads
\begin{equation}
A_1 \equiv \langle \Psi_1| \hat{a}|\Psi_1 \rangle=c_1 c_0=c_1\sqrt{1-c_1^2},
\end{equation}
where $\hat{a}$ denotes the annihilation operator.
After noiseless amplification with gain $g>1$, the state becomes
\begin{equation}
|\tilde{\Psi}_1\rangle =g^{\hat{n}} |\Psi_1\rangle =c_0|0\rangle +g c_1|1\rangle,
\end{equation}
and its amplitude changes to 
\begin{equation}
\tilde{A}_1= \frac{g \sqrt{1-c_1^2} c_1}{1+(g^2-1)c_1^2}.
\end{equation}
The effective amplification gain is given by $\tilde{A}_1/A_1$,
and we get
\begin{equation}
G^{(1)}_{\mathrm{eff}}=\frac{g}{1+(g^2-1)c_1^2}.
\label{Geff1}
\end{equation}
If the probability of single-photon state satisfies $c_1^2>1/(g+1)$, then $G_{\mathrm{eff}}<1$ hence the noiseless
amplification attenuates the complex amplitude of the state. This effect can be understood by noting that the mean amplitude of the superpositions (\ref{psi1})
is maximized when $c_0=c_1=1/\sqrt{2}$. If the amplification gain becomes large enough, then it  enhances the imbalance between
the amplitudes of the vacuum and single-photon contributions, which results in effective reduction of the mean field.
In the limit of very large amplification gain, the noiselessly amplified state approaches a Fock state $|1\rangle$, whose mean field vanishes.

Similar conclusions hold also for the noiseless attenuation. The effective amplitude gain is given again by Eq. (\ref{Geff1}) but with $g=\nu <1$,
\begin{equation}
G^{(1)}_{\mathrm{eff}}=\frac{\nu}{1+(\nu^2-1)c_1^2}.
\end{equation}
If $c_1^2>1/(1+\nu)$ then $G^{(1)}_{\mathrm{eff}}>1$  because starting from a state where the single-photon component is dominant, the noiseless attenuation drives it closer to the balanced superposition
$(|0\rangle+|1\rangle)/\sqrt{2}$.

As a second example, let us consider superposition of the three lowest Fock states,
\begin{equation}
|\Psi_2\rangle=c_0|0\rangle+c_1|1\rangle+c_2|2\rangle,
\end{equation}
where for the sake of simplicity we again assume real $c_j$, and $c_0^2+c_1^2+c_2^2=1$.
The amplitude reads
\begin{equation}
A_2 = \langle \Psi_2|\hat{a}|\Psi_2\rangle =  c_1(c_0+\sqrt{2}c_2).
\end{equation}
Since the formula contains two terms, constructive or destructive quantum interference can occur. After noiseless amplification,
the complex amplitude becomes
\begin{equation}
\tilde{A}_2=\frac{gc_1(c_0+\sqrt{2}g^2c_2)}{c_0^2+g^2c_1^2+g^4c_2^2}
\end{equation}
and the effective amplification gain can be expressed as
\begin{equation}
G^{(2)}_{\mathrm{eff}}=\frac{g}{c_0^2+g^2c_1^2+g^4c_2^2} \times \frac{c_0+\sqrt{2}g^2c_2}{c_0+\sqrt{2}c_2}.
\end{equation}
By suitably choosing $c_0$ and $c_2$, the factor  $c_0+\sqrt{2}g^2c_2$ in the numerator can be made arbitrarily small and we may even achieve zero gain. This can be interpreted as the arising of a destructive interference between the vacuum and two-photon components in the noiselessly amplified state, hence decreasing its mean field.
Similarly, in case of noiseless attenuation, we can choose the parameters such that the factor $c_0+\sqrt{2}c_2$ will be very small and the gain will be very large. Here, the destructive interference that makes the mean field of the initial state very small is disturbed as a result of noiseless attenuation, hence increasing the mean field. Interestingly, this mechanism of interference disturbance is robust against imperfections in the process of noiseless attenuation, so it is a good candidate for an experimental demonstration (see Section V).

We note that the same type of counterintuitive effects may also be exhibited by non-Gaussian mixtures of Gaussian states. Indeed, as a third example, consider the binary mixture of two coherent states $|\alpha\rangle$ and $|\beta\rangle$,
\begin{equation}
\rho_3= p|\alpha\rangle\langle\alpha|+(1-p) |\beta\rangle\langle \beta|,
\label{rhodef}
\end{equation}
where  $p\in[0,1]$. 
The amplitude of this state reads
\begin{equation}
A_3=  \mathrm{Tr}(\rho_3 \hat a) = p\, \alpha+(1-p)\beta.
\end{equation}
After noiseless amplification, each coherent state $|\alpha\rangle$ is mapped onto $|g\alpha\rangle$ with weight factor
$e^{(g^2-1)|\alpha|^2}$. Hence, the resulting state is also a mixture of two coherent states with amplified amplitudes and modified weight,
\begin{equation}
\tilde\rho_3= p'|g\alpha\rangle\langle g\alpha|+(1-p') |g \beta\rangle\langle g \beta|,
\end{equation}
where
\begin{equation}
p'=\frac{p \, e^{(g^2-1)|\alpha|^2}}{ p \, e^{(g^2-1)|\alpha|^2}+(1-p)\, e^{(g^2-1)|\beta|^2}}.
\end{equation}
Its amplitude is given by
\begin{equation}
\tilde A_3=g [p' \alpha + (1-p') \beta]
\end{equation}
so that the effective amplification gain reads
\begin{equation}
G^{(3)}_{\mathrm{eff}}= g \; \frac{p' \alpha + (1-p') \beta}{p\, \alpha+(1-p)\beta}.
\label{Gatt}
\end{equation}
This gain can be complex, and we can have $|G^{(3)}_{\mathrm{eff}}| < 1$ for $g>1$. To see this, take the example of two coherent states with real amplitudes $\alpha=1$ and $\beta= -0.9$ that are mixed with $p=1/3$. If we process this mixture in a noiseless amplifier of gain $g=2$,  we get an effective gain $G^{(3)}_{\mathrm{eff}}=0.063$ smaller than unity. Thus, we observe a mean field reduction by noiseless amplification of a non-Gaussian mixture of coherent states. Conversely, if we set  $g=\nu<1$ in Eq. (\ref{Gatt}),  we get a formula for the effective gain of the noiseless attenuation of state (\ref{rhodef}), and it is easy to find examples where it is larger than $1$. Thus, noiseless attenuation may enhance the mean field amplitude of a non-Gaussian mixture of coherent states.

\begin{figure}[!t!]
\centering
\includegraphics[scale=0.3]{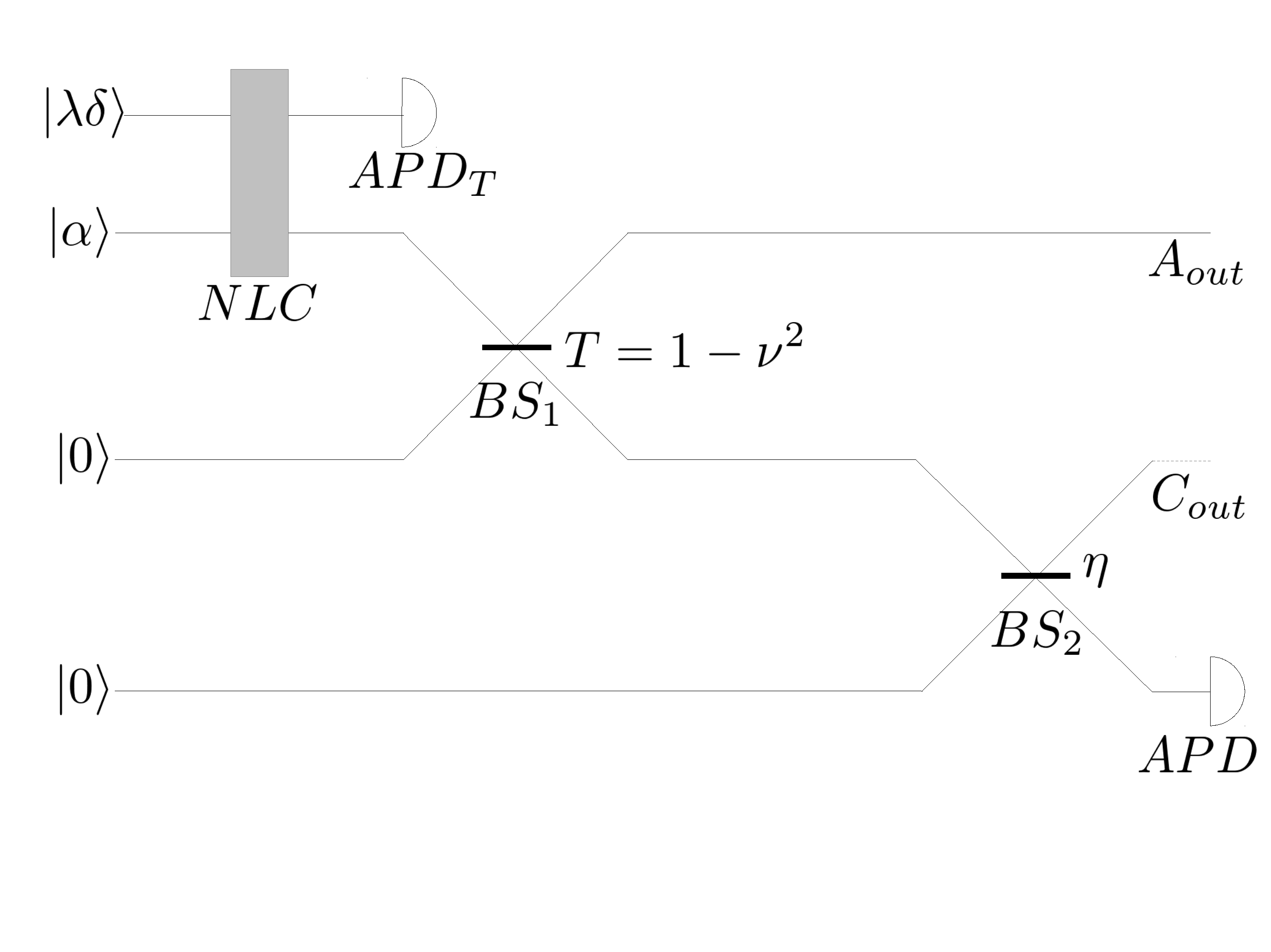}
\caption{Proposed experimental setup. Coherent states are injected into signal and idler ports of a nonlinear crystal where parametric down-conversion
with a low gain $\lambda$ occurs.
Conditional photon addition is heralded by a click of a single-photon detector APD$_T$. BS$_1$ is a beam splitter with amplitude reflectance $\nu$ and
noiseless attenuation is heralded by a no-click of the single-photon detector APD. Imperfect detection with efficiency $\eta <1$, is modeled by coupling to an auxiliary mode C prepared in a vacuum state, where $\eta$ is the transmittance of BS$_2$.}
\label{setup}
\end{figure}

\section{Experimental proposal}

In this section, we propose and analyze an optical setup that enables to experimentally demonstrate the counterintuitive effect of
mean field enhancement by noiseless attenuation. The suggested scheme is illustrated in Fig. \ref{setup}.
The non-Gaussian state is generated from an input coherent state by conditional photon addition. A coherent state $|\alpha\rangle$
is injected into the signal input port of a nonlinear crystal where a non-degenerate parametric down-conversion with a low parametric gain
$\lambda \ll 1$ takes place. A click of the trigger single-photon detector APD$_T$ heralds the generation of a photon pair in the crystal and the addition of
a photon to the signal beam. The noiseless attenuation $\nu^{\hat{n}}$ 
is implemented by sending the signal beam through a beam splitter BS$_1$ with reflectance $R=\nu^2$ and transmittance $T=1-\nu^2$. The auxiliary input port of BS$_1$  is prepared in
vacuum state, and the auxiliary output port is measured with single-photon detector APD. Assuming ideal detector with unit detection efficiency, the noiseless
attenuation is heralded by a no-click of the detector. In practice, the detection efficiency will be rather low, so conditioning on no-clicks will result in a combination
of noiseless attenuation and usual losses. In what follows, we will first assume an ideal APD and then we will provide a more realistic description
which will account for  imperfect state preparation and inefficient single-photon detection.

In order to increase the flexibility of the setup we suggest to also inject a weak auxiliary coherent state $|\lambda\delta\rangle$ to the idler input port of the nonlinear crystal.
The detector APD$_T$ can then be triggered either by the idler photon generated in the crystal or by a photon from the auxiliary input
coherent beam. If these two photons are indistinguishable, then one obtains a coherent superposition of the photon addition and
identity operations, and the resulting conditionally prepared state reads,
\begin{equation}
|\Psi\rangle = \frac{1}{\sqrt{N}}(\hat{a}^\dagger +\delta)|\alpha\rangle.
\label{psiadd}
\end{equation}
Here $N=1+|\alpha^*+\delta|^2$ is a normalization factor and the parameters $\alpha$ and $\delta$ can be independently set to any desired value
by tuning the amplitudes of the coherent beams injected into the signal and idler ports of the nonlinear crystal, respectively. Note that
in the limit $\alpha=0$ the state becomes the superposition of vacuum and single-photon states as studied in the previous section.
For the sake of simplicity, we shall assume that both $\alpha$ and $\delta$ are real. In this case, the complex amplitude
of $|\Psi\rangle$ is real as well,
\begin{equation}
A= \alpha+\frac{\alpha+\delta}{1+(\alpha+\delta)^2}.
\end{equation}
After noiseless attenuation, the state transforms into
\begin{equation}
|\tilde{\Psi}\rangle \propto \nu^{\hat{n}} (\hat{a}^\dagger +\delta) |\alpha\rangle \propto (\nu\hat{a}^\dagger +\delta)|\nu\alpha\rangle.
\label{psiatt}
\end{equation}
where we have used the identity $\nu^{\hat{n}} \hat{a}^\dagger  = \hat{a}^\dagger \nu^{\hat{n}+1}$.
We see that the structure of the state remains unaltered but its parameters change according to $\alpha\rightarrow \nu\alpha$
and $\delta \rightarrow \delta/\nu$. Therefore, the amplitude of the noiselessly attenuated state (\ref{psiatt}) can be expressed as
\begin{equation}
\tilde{A}= \nu\alpha+\frac{\nu\alpha+\delta/\nu}{1+(\nu\alpha+\delta/\nu)^2}.
\end{equation}
The effective gain $G_{\mathrm{eff}}=\tilde{A}/A$ is plotted in Figs.~3 and 4 as a function of $\nu$ and $\delta$, respectively.  
We can see that the effective amplitude gain can be both positive and negative and for suitable parameter values the gain can be much larger than $1$. 
The large gain occurs in the neighborhood of a point where $A=0$, c.f. Fig.~4. 
In the proposed experiment, one could seek an optimal working point $\delta_{\mathrm{opt}}$ such that $G_{\mathrm{eff}}>1$ while the input and output 
amplitudes are large enough so the amplification effect would be observable and not buried in noise.

\begin{figure}[!t!]
\includegraphics[width=0.95\linewidth]{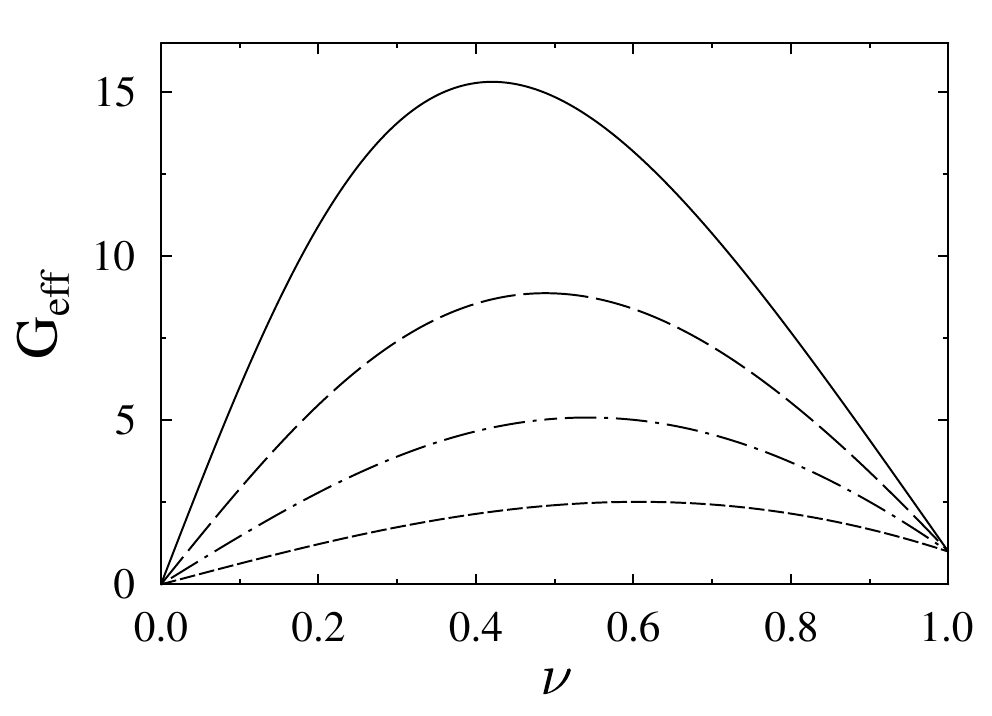}
\caption{Noiseless attenuation of a non-Gaussian state Eq. (\ref{psiadd}). The amplitude gain $G_{\mathrm{eff}}$
is plotted as a function of the attenuation factor $\nu$ for four
different values of detection efficiency $\eta = 1$ (solid line),
$\eta=0.75$ (long dashed line), $\eta=0.5$ (dot dashed line) and
$\eta=0.25$ (short dashed line). The other parameters read $\alpha= 0.25$,
$\delta=-0.55$, and $p=1$.}
\label{Gvsnu02}
\end{figure}

Let us now include the effect of inefficient single-photon detection into our model. As illustrated in Fig. \ref{setup}, this can be done by
including another auxiliary mode C, which is coupled to mode B by the beam splitter BS$_2$ with transmittance $\eta$ equal to the detection
efficiency of the APD. While mode B is projected onto vacuum state, mode C is traced over. The output state before measurement on mode B
reads,
\begin{equation}
\hat{U} (\hat{a}^\dagger+\delta)|\alpha\rangle_A|0\rangle_B|0\rangle_C.
\end{equation}
Here $\hat{U}$ is a unitary operation describing the mode coupling effected by the two beam splitters BS$_1$ and BS$_2$,
\begin{eqnarray}
U=\begin{pmatrix}
  \nu & \sqrt{\eta T} & \sqrt{(1-\eta)T}\\
  -\sqrt{T} & \nu\sqrt{\eta} & \nu\sqrt{1-\eta}\\
  0 & -\sqrt{1-\eta} & \sqrt{\eta}
 \end{pmatrix}&
\label{Unueta}
\end{eqnarray}
where $T=1-\nu^2$.
Hence,
\begin{equation}
\hat{U} \hat{a}^\dagger \hat{U}^\dagger= \nu \hat{a}^\dagger+\sqrt{T}(\sqrt{\eta} \hat{b}^\dagger+\sqrt{1-\eta} \hat{c}^\dagger),
\end{equation}
where $\hat{b}^\dagger$ and $\hat{c}^\dagger$ denote creation operators of modes B and C, respectively, so that the 
output state before measurement on mode B can be rewritten as
\begin{equation}
[\nu \hat{a}^\dagger+\sqrt{T\eta} \, \hat{b}^\dagger+\sqrt{T(1-\eta)} \, \hat{c}^\dagger +\delta]  \;  \hat{U} |\alpha\rangle_A |0\rangle_B|0\rangle_C.
\end{equation}
We also use the fact that a passive linear optical network transforms input 
 coherent states onto output coherent states, so that
\begin{equation}
\hat{U}|\alpha\rangle_A|0\rangle_B|0\rangle_C=|\nu\alpha\rangle_A|\sqrt{T\eta}\alpha\rangle_B|\sqrt{T(1-\eta)}\alpha\rangle_C.
\end{equation}
After projection of mode B onto vacuum, the unnormalized conditional state reads
\begin{equation}
|\tilde{\Psi}_\eta\rangle=(\nu a^\dagger+\sqrt{T(1-\eta)} \, c^\dagger+\delta) \; |\nu\alpha\rangle_A|\sqrt{T(1-\eta)}\alpha\rangle_C.
\end{equation}
The amplitude of the output signal mode A is then given by
\begin{equation}
\tilde{A}_\eta= \nu\alpha +\frac{\nu\alpha(1-\eta T)+\nu\delta}{[\alpha(1-\eta T)+\delta]^2+1-\eta T}.
\end{equation}

\begin{figure}[!b!]
\includegraphics[width=0.95\linewidth]{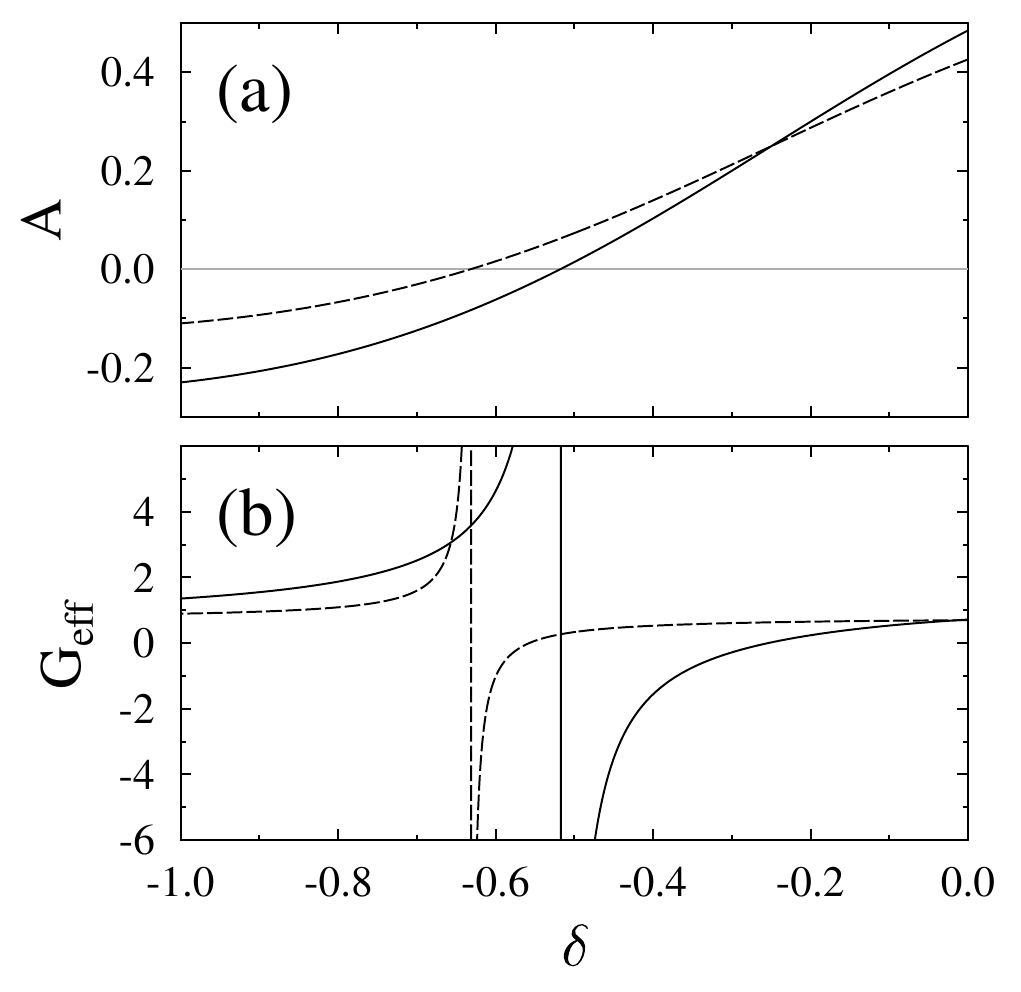}
\caption{Dependence of the input amplitude $A$ (a) and amplitude gain $G_{\mathrm{eff}}$ (b)
on the displacement parameter $\delta$. The other parameters read
$\alpha=0.25$, $\nu=1/\sqrt{2}$, and $\eta=1$, $p=1$ (solid line) and
$\eta=0.25$, $p=0.75$ (dashed line).}
\label{AmpGainvsd}
\end{figure}

A second effect that we take into account is the imperfect mode overlap in conditional photon addition. With some probability, 
the photon may be added to a different mode and in this case the input state remains the coherent state $|\alpha\rangle$.
Thus, a realistic input state can be modeled as a mixture of the state (\ref{psiadd}) and the coherent state $|\alpha\rangle$ ,
\begin{equation}
\rho= p |\Psi\rangle\langle \Psi|+(1-p)|\alpha\rangle\langle \alpha|,
\end{equation}
where $p\in[0,1]$. After noiseless attenuation, the (un-normalized) state of the output signal mode reads
\begin{equation}
\rho_{\mathrm{out}}= \frac{p}{N}\mathrm{Tr}_C[|\tilde{\Psi}_\eta\rangle \langle \tilde{\Psi}_\eta|]+(1-p)|\nu\alpha\rangle\langle \nu\alpha|.
\end{equation}
The amplitude of this output state can be expressed as 
\begin{equation}
\tilde{A}_{\eta,p}= p^\prime\tilde{A}_\eta+(1-p^\prime)\nu\alpha,
\end{equation}
where
\begin{equation}
p^\prime=\frac{p}{p+(1-p)\frac{1+(\alpha+\delta)^2}{[\alpha(1-\eta T)+\delta]^2+1-\eta T}}.
\end{equation}
As shown in Figs.~4 and 5, the effect of amplitude enhancement by noiseless attenuation persists even for $p=0.75$ and low detection efficiency $\eta=0.25$.

\begin{figure}[!t!]
\includegraphics[width=0.8\linewidth]{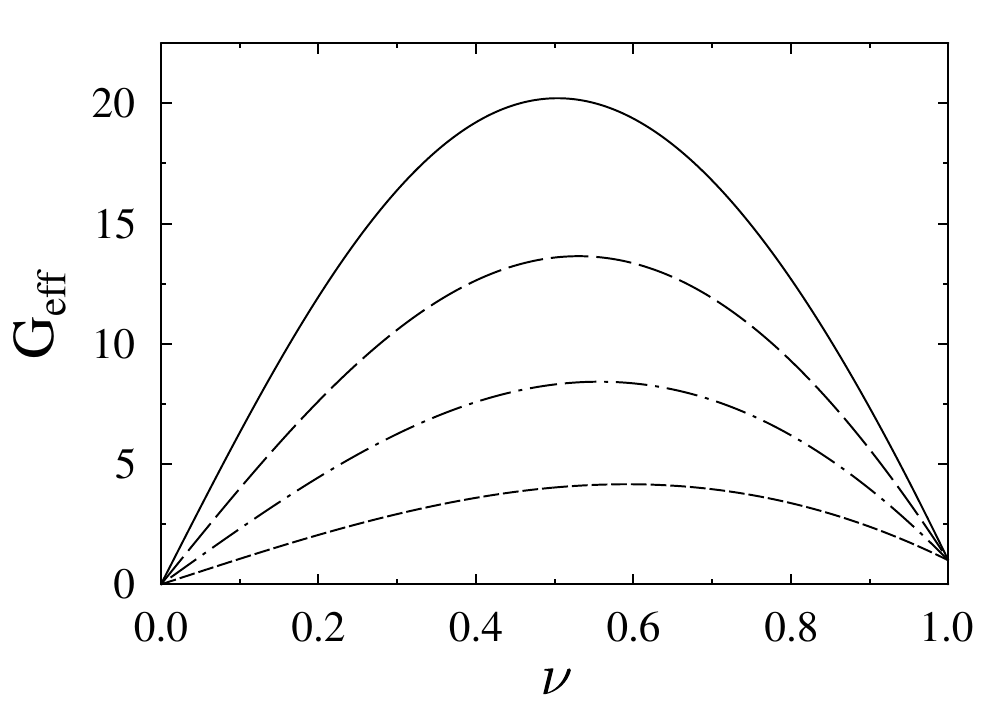}
\caption{The same as Fig. \ref{Gvsnu02} but $p=0.75$ and $\delta=-0.65$.}
\label{Gvsnu03}
\end{figure}

Even severe experimental inefficiencies and imperfections can thus be compensated and give rise to high effective gains by a careful tuning of the amplitude $\delta$. In an experimental realization of the proposed setup, values for the vacuum conditioning efficiency $\eta$ and for the purity parameter $p$ like those used in Figs. 4 and 5 are realistic and probably even too conservative. An efficient vacuum conditioning can be obtained by loosening the spectral and spatial filtering in front of the APD detector so to bring losses in the heralding channel to a minimum. The increased rate of unwanted background counts can be limited by using pulsed laser sources for gating the time interval when the absence of APD clicks should be detected. Note, however, that higher levels of background counts only decrease the no-click heralding rate, without compromising the quality of the generated states. Finally, a regime of small coherent state amplitude $\alpha$ and high reflectance $\nu$ should be preferentially used in an experiment in order to avoid saturation effects in the APD detector.

\section{Conclusions}
We have investigated the behavior of quantum states of light under the action of heralded noiseless attenuation and amplification. By considering certain non-Gaussian states, we have found out that amplifying the state may be accompanied by a decrease of its mean field amplitude $\langle \hat{a} \rangle$. Conversely, noiselessly attenuating the state may come with an increased coherent amplitude.
We have proven that such a counterintuitive effect cannot occur for Gaussian states, so it is specific to non-Gaussian states. 
We have proposed an experimental scheme that is feasible with current technology and should enable the observation of this intriguing phenomenon with a coherently-displaced single-photon added coherent state under realistic experimental conditions. An alternative way to observe this effect may be based on the virtual noiseless amplifier or
attenuator \cite{gausspostselect}, where the amplification or attenuation effect is emulated by post-processing the experimental data.

\acknowledgments
C.N.G. acknowledges financial support from Wallonia-Brussels International via the excellence grants program and gratefully thanks the Department of Optics, Palacky University for the fruitful collaboration and hospitality during his visit in February 2013. This work was supported by the F.R.S.-FNRS under the Eranet project HIPERCOM and by the European Social Fund and Czech Ministry of Education under project EE2.3.20.0060 of Operational Program Education for Competitiveness.
J.F. acknowledges support by the Czech Science Foundation (Project No. P205/12/0577). M.B. and A.Z. acknowledge support from the EU under ERA-NET CHIST-ERA project QSCALE.

\end{document}